%% file: main.tex
\documentclass[%
reprint,
superscriptaddress,
 prl,
 amsmath,
 amssymb,
 aps,
 longbibliography
]{revtex4-2}

\usepackage{cancel}
\usepackage{graphicx}
\usepackage{dcolumn}
\usepackage{bm}
\usepackage[usenames,dvipsnames]{color}
\usepackage{xcolor}
\usepackage{soul}
\usepackage{subcaption}
\usepackage{enumitem}
\usepackage{xspace}
\usepackage{booktabs}
\usepackage{lipsum}
\usepackage{siunitx} 
\usepackage[ISO]{diffcoeff}
\usepackage{subcaption}
\usepackage{ragged2e}
\usepackage{mathrsfs}
\usepackage{hyperref}
\usepackage[capitalise]{cleveref}
\usepackage[normalem]{ulem}
\usepackage{amsmath}
\usepackage{wrapfig}
\usepackage{esvect} 
\usepackage{titlesec}
\usepackage{titletoc}
\usepackage{chngcntr}
\usepackage{fontawesome5}
\usepackage{mleftright}
\usepackage[nodayofweek]{datetime}
\usepackage{comment}

\DeclareSIUnit\year{yr} 
\def\arraystretch{2}\tabcolsep=10pt

\DeclareCaptionJustification{justified}{\justifying}
\definecolor{firebrick}{HTML}{B22222}

\hypersetup{
    colorlinks=true,       
    linkcolor=firebrick,          
    citecolor=firebrick,        
    filecolor=firebrick,      
    urlcolor=firebrick           
}

\definecolor{orcid-green}{RGB} {166, 206, 57}
\newcommand{\MYhref}[3][blue]{\href{#2}{\color{#1}{#3}}}%

\titleclass{\mysection}{straight}[\section]
\titleformat{\mysection}[runin]
  {\itshape}{\thesection}{}{}[.---]
\titlespacing{\mysection}{1em}{1em}{0em}

\newcommand\funop[1]{\mathop{{}#1}}
\newcommand{\dd}{\mathop{}\!\mathrm{d}}
\newcommand{\mrm}[1]{\mathrm{#1}}

\newcommand{\bvec}[1]{\boldsymbol{#1}}

\DeclareSIUnit\clight{c}

\begin{document}

\title{\boldmath First Search for Ultralight Dark Matter Using a Magnetically Levitated Particle}

\author{Dorian W.~P.~Amaral\,\MYhref[orcid-green]{https://orcid.org/0000-0002-1414-932X}{\faOrcid}}
\email{dorian.amaral@rice.edu}
\affiliation{Department of Physics and Astronomy, Rice University, MS-315,
Houston, TX, 77005, U.S.A.}

\author{Dennis G.~Uitenbroek\,\MYhref[orcid-green]{https://orcid.org/0009-0008-3425-6406}{\faOrcid}}
\affiliation{Leiden Institute of Physics, Leiden University, P.O. Box 9504, 2300 RA Leiden, The Netherlands.}

\author{Tjerk H.~Oosterkamp\,\MYhref[orcid-green]{https://orcid.org/0000-0001-6855-5190}{\faOrcid}}
\affiliation{Leiden Institute of Physics, Leiden University, P.O. Box 9504, 2300 RA Leiden, The Netherlands.}

\author{Christopher D.~Tunnell\,\MYhref[orcid-green]{https://orcid.org/0000-0001-8158-7795}{\faOrcid}}
\affiliation{Department of Physics and Astronomy, Rice University, MS-315,
Houston, TX, 77005, U.S.A.}

\begin{abstract}
\noindent
We perform the first search for ultralight dark matter using a magnetically levitated particle. A sub-millimeter permanent magnet is levitated in a superconducting trap with a measured force sensitivity of \SI{0.2}{fN/\sqrt{Hz}}.
   We find no evidence of a signal and derive limits on dark matter coupled to the difference between baryon and lepton number, $B - L$, in the mass range $(1.10360 \text{--} 1.10485) \times 10^{-13}\,\mrm{eV} / c^2$. Our most stringent limit on the coupling strength is $g_{B - L} \lesssim 2.98 \times 10^{-21}$. We propose the POLONAISE (Probing Oscillations using Levitated Objects for Novel Accelerometry In Searches of Exotic physics) experiment, which features short-, medium-, and long-term upgrades that will give us leading sensitivity in a wide mass range, demonstrating the promise of this novel quantum sensing technology in the hunt for dark matter.
\end{abstract}
\maketitle

\mysection{Introduction}
\label{sec:intro}

Dark matter (DM) dominates the matter content of our Universe. Yet, we know remarkably little about its fundamental nature. Except for the fact that it must interact gravitationally, important properties such as its mass, spin, and other potential interactions remain largely a mystery~\cite{Bertone:2016nfn,Freese:2017idy,Cirelli:2024ssz}. 
Astrophysical observations suggest that its mass can lie anywhere in the range of $10^{-19}\,\mrm{eV}/c^2$ to a few solar masses, spanning a vast $90$ orders of magnitude~\cite{Brandt:2016aco,Dalal:2022rmp}.
The lower end of this mass window defines the \textit{ultralight} regime for dark matter, which has been gaining considerable attention~\cite{Essig:2013lka,Antypas:2022asj}.

In this regime, ultralight dark matter (ULDM) particles must be bosonic to reconcile the observed dark matter density. A consequence of this is that these particles exhibit wavelike behavior, leading to interference regions throughout the cosmos. Popular ultralight candidates include the QCD axion~\cite{Weinberg:1977ma,Wilczek:1977pj,DiLuzio:2020wdo}, axion-like particles as well as other scalars~\cite{Arvanitaki:2009fg, Ringwald:2014vqa, Ferreira:2020fam}, and---of relevance to us---vector particles~\cite{Jaeckel:2012mjv,Fabbrichesi:2020wbt,Caputo:2021eaa}.

Ultralight vector dark matter particles are spin-$1$ bosons that stem from the same type of symmetry as the Standard Model photon. Many early-Universe production mechanisms generate this type of dark matter~\cite{Graham:2015rva,Agrawal:2018vin, Co:2018lka,Dror:2018pdh,Bastero-Gil:2018uel,Long:2019lwl,Kolb:2020fwh,Co:2021rhi,Adshead:2023qiw,Cyncynates:2023zwj,Ozsoy:2023gnl}, and the late-Universe structures that it can form have been explored via numerical simulations~\cite{Amin:2022pzv,Gorghetto:2022sue,Jain:2022agt,Jain:2023ojg,Jain:2023qty}. These particles can communicate with us via charges different from that of electromagnetism. In this work, we take this to be the difference between the baryon and lepton numbers of a particle, $B - L$; this leads to a well-motivated dark matter candidate~\cite{Fayet:1980ad,Fayet:1980rr} and can also help to explain the non-zero mass of neutrinos~\cite{Peskin:1995ev,Basso:2008iv, Kanemura:2014rpa}. 

Many experiments have constrained the interaction strength of $B - L$ coupled dark matter, and a variety of detector technologies have been used to derive projected sensitivities. Limits have been set by fifth-force experiments, such as MICROSCOPE~\cite{Touboul:2017grn,Berge:2017ovy,MICROSCOPE:2022doy,Amaral:2024tjg} and Eöt-Wash~\cite{Wagner:2012ui,AxionLimits}, and gravitational wave interferometers, such as LIGO/Virgo~\cite{LIGOScientific:2021ffg} and KAGRA~\cite{LIGOScientific:2024dzy}, among other experiments. Projections with accelerometers have been shown to be promising~\cite{Graham:2015ifn,Carney:2020xol}, with these instruments being realized as torsion balances~\cite{Graham:2015ifn}, optomechanical cavities~\cite{Carney:2019cio,Manley:2020mjq}, atomic interferometers~\cite{abe2021matter}, and future gravitational-wave detectors~\cite{Pierce:2018xmy,Morisaki:2020gui,Nakatsuka:2022gaf,Fedderke:2022ptm}. 

One detector technology undergoing a significant rate of innovation involves the levitation of macroscopic objects via magnetic Meissner levitation~\cite{Vinante2020, Hofer2023, Latorre:2022vxo, Schmidt:2024hdc}. Levitated magnets are excellent force and acceleration sensors~\cite{Fuchs:2023ajk,Janse:2024kcn}, making them ideal for detecting the minuscule signatures expected from ultralight dark matter~\cite{Higgins:2023gwq,Li:2023wcb,Kalia:2024eml,Kilian:2024fsg}.
The low temperatures involved in these setups provide exceptionally low thermal noise and, compared to optical and electrical levitation strategies, much larger levitated objects are possible~\cite{Vinante2020}.
The ability to levitate heavier objects gives us greater sensitivity to dark matter couplings proportional to mass, such as that arising from $B - L$ dark matter~\cite{Kilian:2024fsg}.

In this \textit{Letter}, we perform the first search for ultralight dark matter using a magnetically levitated mass and propose the POLONAISE (Probing Oscillations using Levitated Objects for Novel Accelerometry In Searches of Exotic physics) experiment. We analyze data from the setup initially described in Ref.~\cite{Fuchs:2023ajk} for tests of small-scale gravity, employing a sub-millimetre-sized magnetically levitated particle. Using a likelihood-led treatment to perform our inferencing, we account for the inherent stochasticity in the ULDM field. We motivate {short-,} medium-, and long-term upgrades to propose POLONAISE, which will allow us to achieve leading sensitivity to ULDM and highlight the advancements in quantum metrology necessary for a world-leading DM experiment.

\mysection{Ultralight vector dark matter}
Ultralight vector dark matter consists of  spin-1 bosons with masses $m_\mrm{DM}$ between $10^{-19}\,\mrm{eV}/c^2$ to $2\,\mrm{eV} / c^2$~\cite{Ferreira:2020fam}. To compose all of the local dark matter density, $\rho_\mrm{DM} \approx 0.4\,\mrm{GeV}/c^2\,\mathrm{cm^{-3}}$~\cite{Catena:2009mf,Read:2014qva}, their small mass results in a macroscopic number of particles within a de Broglie volume, $\lambda_\mrm{dB}^3$. Assuming virialization and taking the local circular velocity to be $v_0 \approx \SI{220}{\kilo\meter\per\second}$~\cite{Evans:2018bqy}, we have a total of $N_\mrm{dB} \sim (\rho_\mrm{DM} / m_\mrm{DM}) (h / m_\mrm{DM} v_0)^3 \sim 10^{58}$ particles for a DM mass of $10^{-13}\,\mrm{eV} / c^2$. Consequently, DM particles within the halo behave more as classical waves than particles.

Within a de Broglie volume, these DM waves oscillate coherently at their Compton angular frequency $\omega_\mrm{DM} \equiv 2 \pi f_\mrm{DM} = m_\mrm{DM} c^2 / \hbar$, with a small frequency spread of $\Delta \omega \simeq  (v_0^2 / c^2) \omega_\mrm{DM} \sim 10^{-6} \omega_\mrm{DM}$ between volumes. These define coherent regions that travel at an average velocity of $v_0$, with coherence maintained over the timescale $\tau_\mrm{coh} \equiv \lambda_\mrm{dB} / v_0 = h / (m_\mrm{DM} v_0^2) \approx 21\,\mrm{h}$ for a DM mass of $10^{-13}\,\mrm{eV} / c^2$.
We remain in the coherent regime throughout this work.

Within this regime, the ultralight vector DM field at time $t$, $\bvec{A}(t)$, traces a three-dimensional ellipse. It can be written as~\cite{Amaral:2024tjg,supp_mat}
\begin{equation}
    \bvec{A}(t) = \frac{\hbar}{m_\mrm{DM} c^2}\sqrt{\frac{2 \rho_\mrm{DM}}{3 \varepsilon_0}} \sum_{i}
        \alpha_i \cos(\omega_\mrm{DM} t + \varphi_i)\,\hat{\bvec{e}}_i\,,
\end{equation}
where $\varepsilon_0$ is the permittivity of free space, and the sum runs over the three components of the field, $i \in \{x, y, z\}$. The parameters $\alpha_i$ and $\varphi_i$ respectively control the amplitudes and phases of each of the field's components, and they are stochastic.
Incorporating the randomness in these variables is crucial for accurate inferences, as this can lead to significant correction factors~\cite{Foster:2017hbq,Centers:2019dyn,Lisanti:2021vij,Nakatsuka:2022gaf}.

The ultralight vector dark matter field generates new electric and magnetic fields that can couple to ordinary matter. In the case of $B - L$ dark matter, they arise due to a new interaction term in the Standard Model Lagrangian, $\mathcal{L} \supset - g_{B - L} j^\mu_{B - L} A_\mu$, where $g_{B - L}$ is the gauge coupling strength, $j^\mu_{B - L}$ in the new interaction four-current, and $A^\mu$ is the dark matter four-field~\cite{supp_mat}. For non-relativistic ULDM, the electric field dominates and results in a new Lorentz force, $\bvec{F}(t)$.

\begin{figure}[t!]
    \centering
    \includegraphics[width=\columnwidth]{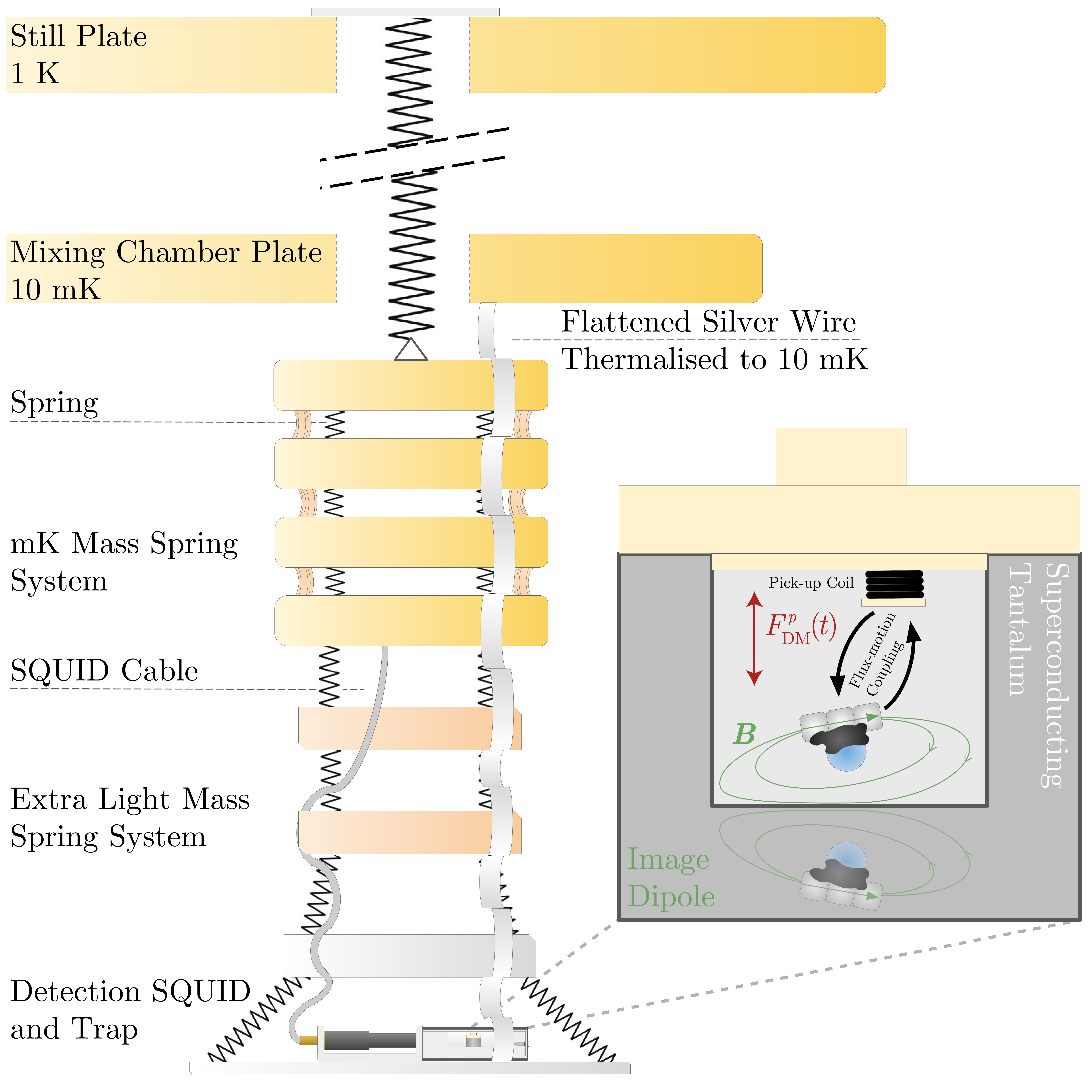}
    \caption{Schematic of the experimental setup inside the dilution refrigerator. Shown are the plates of the cryostat, the multi-stage mass-spring system used to shield against external vibrations, and the holder for the trap and magnet. The inset shows the superconducting trap containing the magnet. The effect of the $\bvec{B}$-field can be modelled by an image dipole. The oscillatory force imparted by the ULDM field, $F_\mrm{DM}^p(t)$, is also indicated. The black arrows illustrate the coupling of the flux of the moving particle to the pick-up coil. Further details and photographs of the setup can be found in Ref.~\cite{Fuchs:2023ajk}.}
    \label{fig:exp}
\end{figure}

Our experiment consists of a magnetically levitated particle suspended within a superconducting trap, illustrated in \cref{fig:exp} and described in detail below. We observe the motion of the particle for a time $T_\mrm{obs}$, recording it along a single sensitivity axis $\bvec{\zeta}$ via a superconducting pick-up coil. This makes the relevant force $F_\mrm{DM}(t) \equiv \bvec{\zeta} \cdot \bvec{F}(t)$, where we have approximated $\bvec{\zeta}$ to be static since our observation time is shorter than a day. With the current level of vibration isolation, our experiment is most sensitive along the zenith; we thus align $\zeta$ in this direction. Using all three translational degrees of motion would allow us to measure the polarization of the ULDM field in the future.

Accounting for the dynamics of both the trap and the levitated particle, the force experienced by the latter is
\begin{equation}
\begin{split}
F^p_\mrm{DM}(t) \simeq \mathcal{F}[&\alpha_x \cos\lambda \cos\phi \cos(\omega_\mrm{DM}t + \varphi_x) \\
+\, &\alpha_y \cos\lambda \sin\phi \cos(\omega_\mrm{DM}t + \varphi_y) \\
+\, &\alpha_z \sin\lambda \cos(\omega_\mrm{DM}t + \varphi_z)]\,,
\end{split}
\label{eq:dm-force-signal}
\end{equation}
where we have defined the force scale
\begin{equation}
\mathcal{F} \equiv g_{B - L} \left(\mathcal{R}_p - \frac{\omega_0^2}{\omega_\mathrm{DM}^2}\mathcal{R}_t \right)  m_p a_0\,,
\label{eq:force-amp}
\end{equation}
with $\mathcal{R}_p$ and $\mathcal{R}_t$ the averaged neutron-to-atomic-weight ratios of the particle and trap, respectively, $\omega_0$ the resonance angular frequency of the particle, $m_p$ the total mass of the particle, and $a_0 \approx 2.12 \times 10^{11}\,\mrm{m\,s^{-2}}$ a characteristic acceleration imparted by the ULDM field~\cite{supp_mat}.

We perform inferences in Fourier space by analyzing the force power spectral density (PSD). The PSD will contain the power delivered to our setup by external sources, as well as that supplied by a ULDM signal.  When observing for less than a day, the signal is a monochromatic peak in the PSD within the bin containing the Compton frequency. The quantity of interest for our inferencing is the \textit{excess power} at that frequency, defined as the value of the PSD normalized by the expected noise, $S_{FF}$, within that bin. The excess power is proportional to the square of the dimensionless parameter~\cite{supp_mat}
\begin{equation}
    \kappa \equiv \sqrt{\frac{\mathcal{F}^2 T_\mathrm{obs}}{2 S_{FF}}}\,,
    \label{eq:beta}
\end{equation}
By measuring the force PSD and fitting to our expected noise level, we perform inferences on $\kappa$ and map these values to those for the coupling strength, $g_{B - L}$:
\begin{equation}
g_{B - L} = \frac{\kappa}{\left\lvert\mathcal{R}_p - (\omega_0^2/\omega_\mathrm{DM}^2)\mathcal{R}_t\right\rvert m_p a_0 }\sqrt{\frac{2 S_{FF}}{T_\mathrm{obs}}}\,.
\label{eq:gBL}
\end{equation}
From this, it is clear that magnetic levitation, which affords us the use of heavier particle masses, is an ideal levitation strategy for a bosonic ULDM search.

\mysection{Experiment}

Originally designed to detect small-scale gravity~\cite{Fuchs:2023ajk}, our setup has the force sensitivity and frequency range required of a promising ULDM detector.
It features a Type-I superconducting trap with a magnetically levitated permanent magnet composed of three \SI{0.25}{mm} Nd$_{2}$Fe$_{14}$B cubes and a spherical glass bead of radius \SI{0.25}{mm} to break rotational symmetry (Fig.~\ref{fig:exp}). 
The levitated particle has a mass of $m_p \approx \SI{0.43}{mg}$, a resonance frequency of $f_0 \approx 26.7\,\mrm{Hz}$ for motions along the zenith, and a quality factor of $Q \approx 9.3 \times 10^6$~\cite{Fuchs:2023ajk}. We calculate $\mathcal{R}_p \approx 0.518$~\cite{supp_mat}. 

We detect the motion of the particle using a superconducting pick-up loop. The motion of the magnet induces a change in flux in the loop, causing a superconducting current to run in the circuit. This circuit consists of the pick-up loop, the calibration loop, and the SQUID input coil, which is inductively coupled to a two-stage direct current SQUID. The calibration loop calibrates the energy coupling between the detection circuit and the degrees of motion of the magnet. We find $\mathcal{R}_t \approx 0.526$~\cite{supp_mat}. 

\begin{figure}
    \centering
\includegraphics{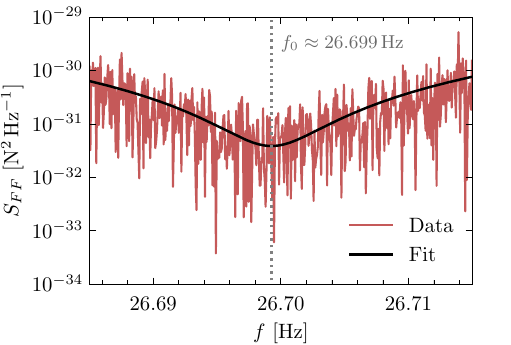}
    \caption{The force-power spectral density, $S_{FF}$, with frequency $f$ measured by our experiment. The fit to the force noise background via \cref{eq:force-noise-bkg} is shown. The vertical line highlights the resonance frequency $f_0 \approx 26.699\,\mrm{Hz}$ relevant for motions parallel to the zenith. Data initially reported in Ref.~\cite{Fuchs:2023ajk}.}
    \label{fig:exp-data}
\end{figure}

We shield the trap from environmental vibrations both vertically and laterally using a multi-stage mass-spring system, and we thermalize the experiment via a flexible silver wire. The closest part of this system, consisting of seven masses, hangs from the still plate, which is suspended from the \SI{3}{K} plate of the cryostat.
The cryostat rests on a $25$-ton concrete block with pneumatic dampers to reduce vibrations caused by the building. The pulse tube cooler and vacuum pumps are mounted separately.

The force noise data gathered in Ref.~\cite{Fuchs:2023ajk} is shown in \cref{fig:exp-data}. The data was continuously recorded on \formatdate{13}{05}{2022} over a time of $T_\mrm{obs} \approx 4.2\,\mrm{h}$ between \formattime{22}{32}{00} and \formattime{02}{44}{00} UTC and spans the frequency range $26.6850\,\mrm{Hz}\text{--}26.7150\,\mrm{Hz}$. 
Compared to the original data, we have removed two points: one at the frequency of a generated monochromatic signal (a spinning wheel used for a gravity test) and another at the resonance frequency. The latter was done as the ringdown time of the resonator exceeded the dataset length, so this point represented the starting amplitude of the resonator (see supplement E of Ref.~\cite{Fuchs:2023ajk}).
We fit the data to the total expected force noise via
\begin{equation}
    \label{eq:force-noise-bkg}
    S_{FF}(\omega) = S_{FF}^0 + |\chi(\omega)|^{-2} S_{xx}\,,
\end{equation}
where $S_{FF}^0$ is the force noise arising from the position measurement of the particle, and  $S_{xx}$ is the displacement noise due to measurement-added noise from the SQUID. The factor $\chi(\omega) \equiv [m_p(\omega_0^2 - \omega^2 + i \gamma \omega)]^{-1}$ is the mechanical susceptibility, with $\omega_0 \equiv 2 \pi f_0$ and $\gamma \equiv \omega_0 / Q$. We find the fit values $S_{FF}^0 \approx 3.88 \times 10^{-32}\,\mrm{N^2\,Hz^{-1}}$ and $S_{xx} \approx 3.59 \times 10^{-21}\,\mrm{m^2\,Hz^{-1}}$.

\mysection{Data analysis and results}

The ULDM signal manifests as a monochromatic peak in the force PSD within the bin containing the Compton frequency, $f_\mrm{DM}$. 
Its location is dictated by the DM mass, and its noise-normalized amplitude is determined by the parameter $\kappa$. We scan over our frequency bins, equivalent to scanning over dark matter masses $m_\mrm{DM}$, and use our measured force PSD to make inferences on the coupling strength, $g_{B - L}$.

To perform our search, we follow a frequentist approach similar to Ref.~\cite{Amaral:2024tjg}. The likelihood of measuring a value for the excess power in any frequency bin is a non-central $\chi^2$ with two degrees of freedom and a non-centrality parameter controlled by $\kappa$. To account for the stochasticity of the ULDM field, we marginalize this likelihood over the three random Rayleigh amplitudes and uniform phases. The result is an exponential likelihood with inverse scale $2 + \kappa^2$~\cite{supp_mat}.

We use this likelihood to define a two-sided test statistic (TS) based on the log-likelihood ratio to ascertain whether a DM signal is present. As our discovery criterion, we require that the excess power yields at least a $3\sigma$ significance over the expected background, equivalent to a $p$-value of  $2.7 \times 10^{-3}$. We evaluate this by first building the distribution of the TS under the null hypothesis that no such signal exists using toy Monte Carlo (MC) simulations. At every frequency bin, we then compute the $p$-value of the data. Our most significant local $p$-value is $p \approx 2.0 \times 10^{-2}$, which does not exceed our discovery threshold. We therefore turn to setting a limit.

We place $90\%$ confidence level (CL) limits on $g_{B - L}$ using a similar procedure, but taking the signal hypothesis as our null hypothesis.
To assess whether our limit is consistent with the background model, we derive the expected median limit and $1\sigma$/ $2\sigma$ limit bands from our setup by simulating background-only pseudodata. We find $\kappa_\mrm{lim}^\mrm{med} \approx 3.85$, with $1\sigma$ and $2\sigma$ bands given by $\kappa_\mrm{lim}^{1\sigma} \in [2.01, 6.47]$ and $\kappa_\mrm{lim}^{2\sigma} \in [1.23, 9.39]$. Further details on our statistical procedure can be found in Ref.~\cite{supp_mat}.

\begin{figure}
    \centering
\includegraphics{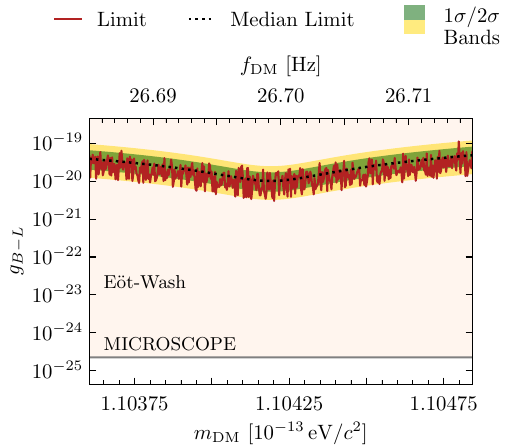}
    \caption{The $90\%$ confidence level limits on the gauge coupling strength $g_{B - L}$ with dark matter mass $m_\mrm{DM}$ (bottom axis) and Compton frequency $f_\mrm{DM}$ (top axis). Shown is the data-driven limit derived from the measurements presented in \cref{fig:exp-data}, as well as the median limit and $1\sigma$/$2\sigma$ bands derived from our Monte Carlo analysis \cite{supp_mat}. Also shown are the existing limits from the Eöt-Wash~\cite{Wagner:2012ui,AxionLimits} and MICROSCOPE~\cite{MICROSCOPE:2022doy,Amaral:2024tjg} experiments.}
    \label{fig:present-lims}
\end{figure}

We show our $90\%$ CL limit in \cref{fig:present-lims}. We find excellent agreement between the results from our MC analysis and our derived limit, which closely follows the median limit and lies well within the $2\sigma$ band. Our best constraint is $g_{B - L} \lesssim 2.98 \times 10^{-21}$, occurring at the DM mass $m_\mrm{DM} \approx 1.1042 \times 10^{-13}\,\mrm{eV} / c^2$ (Compton frequency $f_\mrm{DM} \approx 26.6995\,\mrm{Hz}$). This limit is not as stringent as those set by the fifth-force Eöt-Wash~\cite{Wagner:2012ui,AxionLimits} and MICROSCOPE~\cite{MICROSCOPE:2022doy,Amaral:2024tjg} experiments, falling at $g_{B - L} \lesssim 10^{-23}$ and $g_{B - L} \lesssim 2 \times 10^{-25}$, respectively. 

Our limit is the first data-driven constraint on ultralight dark matter using a magnetically levitated particle. However, we derived it using a setup designed for tests of gravity---an orthogonal research objective~\cite{Fuchs:2023ajk}. We offer a set of experimentally driven upgrades that will make this detector technology a leading option in the search for dark matter, proposing the POLONAISE experiment.

\mysection{Future Prospects}
\label{sec:beyond}

We propose the first optimization of a magnetically-levitated setup to achieve leading ULDM sensitivities with POLONAISE. Our key improvement is adding a second coil to control the resonant frequency, allowing us to probe a wider DM mass window. To maximize sensitivity, we aim to reduce force noise, use heavier levitated masses, and increase the neutron-to-atomic-weight ratio difference between the particle and trap. We plan to perform a two-year-long resonant scan with short-, medium-, and long-term upgrades as summarized in \cref{tab:configs}.  We estimate these upgrades will take $3$, $5$, and $10$ years, respectively. Details on our choices and their impact on our sensitivity can be found in Ref.~\cite{supp_mat}. 

\begin{table}[t!]
\renewcommand{\arraystretch}{1.5}
    \centering
    \begin{tabular*}{\columnwidth}{@{\extracolsep{\fill}}lccc}
    \toprule
    \midrule
     & Short & Medium & Long\\
    \midrule 
     $T~\mrm{[mK]}$ & $20$ & $20$ & $2$ \\
     $m_p~\mrm{[mg]}$ & $0.43$ & $430$ & $430$ \\
     $n_\mrm{SQ}~\mrm{[\hbar]}$ & $10^3$ & 
     $10^2$ & $10^1$ \\
     $\sqrt{S_{FF}}~\mrm{[N / \sqrt{Hz}]}$ & $10^{-19} f_0^{1/2}$ & $10^{-18}f_0^{1/2}$ & $10^{-19} f_0^{1/2}$ \\
     $\Delta f_\mrm{opt}~\mrm{[mHz]}$ & $3.4$ & $3.4$ & $0.34$ \\
     $Q$ & $10^8$ & $10^9$ & $10^{10}$ \\
     $\left\lvert\mathcal{R}_p - \mathcal{R}_t\right\rvert$ & 0.039 & 0.039 & 0.213 \\
     $N_p$ & $1$ & $10$ & $100$ \\
     \midrule
     \bottomrule    
    \end{tabular*}
    \caption{The short-, medium-, and long-term upgrades for POLONAISE. Shown are the cooling temperature ($T$), the mass of the levitated particle $(m_p)$, the SQUID energy resolution ($n_\mrm{SQ}$), the total root force PSD ($\sqrt{S_{FF}}$), the bandwidth that optimizes the expected force noise ($\Delta f_\mrm{opt}$), the quality factor ($Q$), the difference between the neutron-to-atomic-weight ratios of the levitated particle and trap ($\lvert \mathcal{R}_p - \mathcal{R}_t \lvert$), and the number of levitated particles employed ($N_p$). At each resonance frequency, $f_0$, the measurement time is $T_\mrm{obs} = 4.05 \times 10^5 / f_0$.}
    \label{tab:configs}
\end{table}

At each resonance frequency, $f_0$, we will measure for $4.05 \times 10^5$ Compton cycles to ensure coherence, giving $T_\mrm{obs} = 4.05 \times 10^5 / f_0$. We will match our frequency steps to the optimal bandwidth, $\Delta f_\mrm{opt}$, that maximizes our sensitivity while minimizing noise by tuning the SQUID coupling to have equal backaction and thermal noise.  Cooling will reach \SI{20}{mK} (short/medium-term) and \SI{2}{mK} (long-term) using nuclear demagnetization~\cite{vanHeck:2022evk}. We will improve the quality factor by using insulating magnets to reduce Eddy current damping~\cite{Timberlake:2019swe} and improve SQUID energy resolution with alternating current readouts~\cite{muck2001superconducting}.  Avoiding interference in the SQUID readout will reduce $n_{\mrm{SQ}}$ by factors of $10$ (short-term) and $100$-$1000$ (medium/long-term) with radio frequency readouts~\cite{Schmidt:2024hdc}. 
We will parallelize our scan by using multiple levitated particles monitored by a single SQUID, which can read out multiple pick-up coils in series by designing the traps to have non-overlapping resonance frequencies.
The number of SQUIDS is scalable, with a commercial dilution refrigerator able to house up to $100$ SQUIDs. In the short term, we will scan a range of $21\,\mrm{Hz}\text{--}35\,\mrm{Hz}$, extending this to $10\,\mrm{Hz}\text{--}200\,\mrm{Hz}$ in the medium and long terms. This scan is realizable by adding a coil inside the superconducting trap and using a DC current to tune the resonance frequencies. Using a persistent current will minimize current-added noise~\cite{vanWaarde:2016kud}.

Exploiting magnetic levitation's ability to support heavier objects, we will improve our sensitivity by increasing the particle mass to $43\,\mrm{mg}$ (short-term) and $430\,\mrm{mg}$ (medium/long-term). 
Maintaining force noise and reducing trap vibrations will be challenging as the spring constant increases with mass, requiring vibration isolation improvements of \SI{40}{dB}, \SI{80}{dB}, and \SI{100}{dB} at $25\,\mrm{Hz}$ for each of our upgrades, respectively. This is feasible since the required isolation is comparable to LIGO's performance but only needed around the particle's resonance frequency~\cite{Maggiore:2019uih}.
Finally, by optimizing the experiment materials, we will improve $\lvert \mathcal{R}_p - \mathcal{R}_t\lvert$ to $0.039$ (short/medium-term) and $0.213$ (long-term)~\cite{supp_mat}.
For our long-term goal, we will attach a thin-walled aluminium container filled with 1 liter of solid hydrogen.

\begin{figure}[t!]
    \centering
\includegraphics{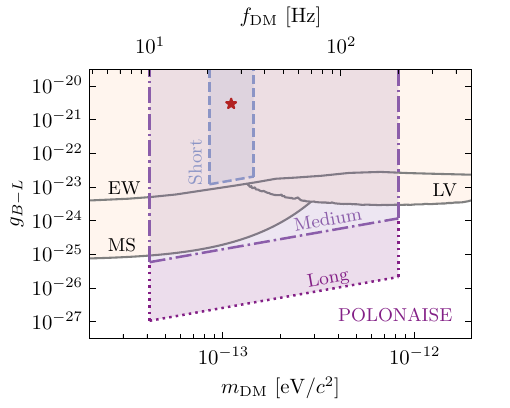}
    \caption{Projected $90\%$ confidence level limits on the gauge coupling strength $g_{B - L}$ with dark matter mass $m_\mrm{DM}$ (bottom axis) and Compton frequency $f_\mrm{DM}$ (top axis) for POLONAISE. Projections are based on the short-, medium-, and long-term configurations detailed in \cref{tab:configs}. Also shown is our current best limit from \cref{fig:present-lims} (red star), as well as the existing limits from the MICROSCOPE (MS)~\cite{MICROSCOPE:2022doy,Amaral:2024tjg}, Eöt-Wash (EW)~\cite{Wagner:2012ui,AxionLimits}, and LIGO/Virgo (LV)~\cite{LIGOScientific:2021ffg,AxionLimits} experiments.}
    \label{fig:proj-lims}
\end{figure}

We show the projected $90\%$ CL limits on the coupling strength from our proposed upgrades in \cref{fig:proj-lims}. To compute them, we have used our derived value of $\kappa_\mrm{lim}^\mrm{med} \approx 3.85$ in \cref{eq:gBL}, giving us projections assuming a background-only measurement. Our short-term upgrade is competitive with Eöt-Wash~\cite{Wagner:2012ui,AxionLimits}, and our medium- and long-term configurations probe new areas of parameter space. Our medium-term limit surpasses that of Eöt-Wash for all scanned masses, is competitive with MICROSCOPE at low masses, and constrains new couplings beyond those of MICROSCOPE and LIGO/Virgo~\cite{LIGOScientific:2021ffg,AxionLimits} at high masses. Our long-term limit is leading, going beyond the limit set by MICROSCOPE~\cite{MICROSCOPE:2022doy,Amaral:2024tjg} by two orders of magnitude at lower masses and that set by LIGO/Virgo by three orders of magnitude at higher masses. 

\mysection{Conclusions}
\label{sec:conc}

We have performed the first search for ultralight dark matter using a magnetically levitated particle. Focusing on dark matter coupled to the difference between baryon and lepton number, $B - L$, we have found no significant evidence of a signal in the mass range $(1.10360 \text{--} 1.10485) \times 10^{-13}\,\mrm{eV} / c^2$, equivalent to the frequency range $26.6850\,\mrm{Hz}\text{--}26.7150\,\mrm{Hz}$. We have set a $90\%$ confidence level limit on the coupling strength, finding the best constraint to be $g_{B - L} \lesssim 2.98 \times 10^{-21}$. We have proposed the POLONAISE experiment, featuring short-, medium-, and long-term improvements to our setup to enable sensitivity to unexplored model parameter space. Our result highlights the promise of this quantum sensing technology in the hunt for dark matter, and we hope that it fuels initiatives in advancing experimental designs of magnetically levitated setups for astroparticle physics.

\acknowledgments
We would like to thank Daniel Carney, Andrew Long, Bas Hensen, Mudit Jain, Martijn Janse, and Yue Zhao for their valuable feedback on the manuscript. DA would like to thank Juehang Qin for helpful discussions throughout this work, especially concerning our statistical treatment, as well as Mustafa Amin for discussions regarding the ultralight dark matter field. DA and CT were jointly funded by Rice University and NSF award 2046549. TO acknowledges funding from NWO grant OCENW.GROOT.2019.088. DU and TO were jointly funded by the EU Horizon Europe EIC Pathfinder project QuCoM (10032223). 

\nocite{Chaudhuri:2014dla,Cowan:2010js,Fedderke:2021aqo,Graham:2014sha,groth1975probability}
\bibliography{biblio}

\clearpage

\appendix

\input{appendices/Appendix_theory}

\input{appendices/Appendix_force}

\input{appendices/Appendix_materials}

\input{appendices/Appendix_stats}

\input{appendices/Appendix_future_experiments}
\end{document}

%% file: appendices/Appendix_theory.tex
\onecolumngrid

\begin{center}
\large
\textbf{
    \textit{Supplemental Material:} \\ First Search for Ultralight Dark Matter Using a Magnetically Levitated Particle
    }
    
\vspace{1.75ex}

Dorian W.~P.~Amaral, Dennis G.~Uitenbroek, Tjerk H.~Oosterkamp, and Christopher D.~Tunnell

\vspace{2ex}

\end{center}

\twocolumngrid

\section{A.~~Ultralight $B - L$ Dark Matter}

\noindent
The Lagrangian for a new, massive gauge field that couples to the difference between the baryon and lepton number, $B - L$, reads
\begin{equation}
\begin{split}
    \mathcal{L} \supset &-
    \frac{\varepsilon_0 c^2}{4} A^{\mu \nu}A_{\mu \nu} + \frac{\varepsilon_0}{2} \left(\frac{m_{A'} c^2}{\hbar}\right)^2 A^{\mu}A_{\mu} \\ 
    &- g_{B - L} j_{B - L}^\mu A_\mu\,,
    \end{split}
    \label{eq:lagrangian}
\end{equation}
where $A^{\mu\nu} \equiv \partial^\mu A^\nu - \partial^\nu A^\mu$ is the field strength tensor for the new field, $m_{A}$ is the mass of the field, and $j^\mu_{B-L}$ is the $B-L$ vector current. Since we take this field to be dark matter, we equate $m_{A} = m_\mrm{DM}$. The vector current is given by
\begin{equation}
    j^\mu_{B-L} = \sum_f Q_{B - L}^f \bar{f} \gamma^\mu f\,,
\end{equation}
where the sum runs over all fermions in the Standard Model, and $Q_{B - L}^f$ is the $B - L$ charge of  fermion $f$. 

In the ultralight regime, we can write the dark matter field at time $t$ and position $\bvec{x}$ as~\cite{Amaral:2024tjg}
\begin{equation}
\begin{split}
   \bvec{A}(\bvec{x}, t) &= |\bvec{A}| \mrm{Re}\mleft\{\frac{1}{\sqrt{V}} \sum_{\bvec{k}} \sqrt{f_{\bvec{k}}} 
   e^{-i( \omega_\mrm{DM}^{\bvec{k}} t + \bvec{k} \cdot \bvec{x})} \bvec{\epsilon}_{\bvec{k}}\mright\}\,,
\end{split}
   \label{eq:full-field}
\end{equation}
where the sum runs over the possible wavevectors of the field, $\bvec{k} = m_\mrm{DM} \bvec{v} / \hbar$ (which are dictated by the Milky Way halo distribution $f_{\bvec{k}}$),  the angular velocity of the field is given by $\omega_\mrm{DM}^{\bvec{k}} \equiv m_\mrm{DM}c^2 / \hbar + \hbar^2|\bvec{k}|^2 / (2 m_\mrm{DM})$, and $\bvec{\epsilon}_{\bvec{k}}$ is a random variable encoding the phases of the field at each wavevector. The volume factor $V^{-1/2}$ appears as we have discretized the field. The halo distribution function is given by
\begin{equation}
    f_{\bvec{k}} \equiv \mathcal{N} \exp\left(-\frac{|\bvec{k} - \bar{\bvec{k}}|^2}{2 k_0^2}\right) \quad \text{with} \quad \frac{\mathcal{N}}{V} \sum_{\bvec{k}} f_{\bvec{k}} = 1\,,
\end{equation}
where $k_0 \equiv m_\mrm{DM} v_0 / \hbar$ is the wavevector of the field at the virial velocity $v_0$.

In the coherent regime, where the observation time is much shorter than the dark matter coherence time $\tau_\mrm{coh} \equiv \lambda_\mrm{dB} / v_0 = h / (m_\mrm{DM} v_0^2)$, we can neglect the kinetic term in $\omega_\mrm{DM}^{\bvec{k}}$, such that $\omega_\mrm{DM}^{\bvec{k}} \simeq \omega_\mrm{DM} \equiv m_\mrm{DM} c^2 / \hbar$. Moreover, since spatial variations in the field occur at the scale $\lambda_\mrm{dB}$, which is smaller than either the separation between our trap and levitated particle or the distance the Earth covers in its orbit during our measurement time, we set $\bvec{x} = \bvec{0}$ assuming statistical homogeneity of the field. This simplifies the expression for the field in \cref{eq:full-field} to    
\begin{equation}
    \bvec{A}(t) \simeq |\bvec{A}| \sum_{i} \alpha_i \cos(\omega_\mrm{DM} t + \varphi_i)\, \hat{\bvec{e}}_i\,,
    \label{eq:coh-field}
\end{equation}
where the sum runs over $i \in \{x, y, z\}$,  $\hat{\bvec{e}}_i$ are the Cartesian unit basis vectors, and where the quantities $\alpha_i$ and $\varphi_i$ are random variables describing the amplitude and phase of the dark matter field, respectively. The three independent $\alpha_i$ follow Rayleigh distributions with variance $1/2$, and the three independent phases $\varphi_i$ follow uniform distributions from $0$ to $2\pi$:
\begin{equation}
        p(\alpha_i) = 2 \alpha_i e^{-\alpha_i^2} \qquad \text{and} \qquad
        p(\varphi_i) = \frac{1}{2 \pi}\,.
        \label{eq:dm-dist}
\end{equation}
Further details and derivations of the above can be found in Ref.~\cite{Amaral:2024tjg}.

Equating the energy density of the field to that of dark matter sets the normalization of the field. In the non-relativistic regime, we have that
\begin{equation}
    \frac{\varepsilon_0}{2} \left[\langle|\dot{\bvec{A}}|^2\rangle +  \left(\frac{m_\mrm{DM} c^2}{\hbar}\right)^2 \langle|\bvec{A}|^2\rangle\right] = \rho_\mrm{DM}\,.
\end{equation}
From \cref{eq:coh-field} and \cref{eq:dm-dist}, we can compute all relevant expectation values. This gives us
\begin{equation}
    |\bvec{A}| = \frac{\hbar}{m_\mrm{DM} c^2}\sqrt{\frac{2 \rho_\mrm{DM}}{3 \varepsilon_0}}\,,
\end{equation}

To make contact with our experiment, we can derive the force that the new gauge field imparts on any object charged under it. This force will take the form of a new Lorentz force created by the electric and magnetic fields sourced by the new gauge field, $\bvec{E}$ and $\bvec{B}$ respectively. Writing $A^\mu \equiv (\phi / c,\, \bvec{A})$, where $\phi$ and $\bvec{A'}$ are respectively the new scalar and vector potentials, we have that
\begin{equation}
    \bvec{E} = -\bvec{\nabla} \phi - \diffp{\bvec{A}}{t} \qquad \text{and} \qquad \bvec{B} = \bvec{\nabla} \times \bvec{A}\,.
\end{equation}
While the temporal derivative picks out a factor of $\omega_\mrm{DM} \propto c^2$, the spatial derivatives lead to factors of $|\bvec{k}| \propto v_0$. Since $v_0 \sim 10^{-3} c$, terms involving spatial derivatives are suppressed compared to the temporal derivative term, such that the dark magnetic field is negligible. Moreover, in the Lorenz gauge ($\partial_\mu A^\mu = 0$), the scalar potential is also suppressed compared to the vector potential by a factor of $v_0$. Thus, we only need to consider the electric field produced by the dark matter field. Given a macroscopic, neutral object of total mass $m$, its net $B - L$ charge will be the total number of neutrons within it, $\mathcal{N}$. We can therefore write the total force that the ultralight DM field should impart on an object as
\begin{equation}
\begin{split}
    \bvec{F}(t) &\simeq g_{B - L} \mathcal{N} \bvec{E} \\
    &\simeq g_{B - L} \left(\frac{\mathcal{N}}{\mathcal{A}}\right) \left(\frac{\mathcal{A}u}{u}\right) \bvec{E} \\
    &\simeq g_{B - L} \mathcal{R} m a_0 \sum_{i}  \alpha_i \cos(\omega_\mathrm{DM} t + \varphi_i)\,\bvec{e}_i\,,
    \label{eq:dm-force}
\end{split}
\end{equation}
where we have absorbed an irrelevant phase of $\pi/2$ into each $\varphi_i$. We have also defined the ratio of neutrons to atomic mass as $\mathcal{R} \equiv \mathcal{N} / \mathcal{A}$ (with $\mathcal{A}$ the atomic mass of the object), approximated the mass of the object as $m \simeq \mathcal{A} u$ (with $u \approx 1.66 \times 10^{-27}\,\mrm{kg}$ the atomic mass unit), and introduced the acceleration given per nucleon as
\begin{equation}
    a_0 \equiv \frac{1}{u} \sqrt{\frac{2 e^2 \rho_\mrm{DM}}{3 \varepsilon_0}} \approx 2.12 \times 10^{11}\,\mrm{m\,s^{-2}}\,,
\end{equation}
where the DM energy density is $\rho_\mrm{DM} \approx 0.4\,\mrm{GeV}/c^2\, \mrm{cm^{-3}}$~\cite{Read:2014qva}.

We note that, while the focus of this work is on ultralight vector dark matter, magnetically levitated setups such as ours have also been shown to have excellent sensitivity to other ultralight dark matter candidates. In particular, our experiment should also have sensitivity to axion-like particles and dark photon dark matter~\cite{Higgins:2023gwq,Kalia:2024eml,Kilian:2024fsg}.

Both of these dark matter candidates would leave their signatures as an alternating magnetic field within the superconducting trap~\cite{Chaudhuri:2014dla, Fedderke:2021aqo, Graham:2014sha, Kalia:2024eml}. This differs from our case, where it is the induced electric field that causes the detectable effect. The magnetic field would result in a torque on our levitated particle, measurable by the pick-up coil installed in our trap. For axion-like particles, this would allow us to probe the axion-electron and axion-photon couplings. For dark photon dark matter, following a displacement of our particle from the centre of the trap, we would have sensitivity to the kinetic mixing parameter between the Standard Model photon and the dark photon~\cite{Kalia:2024eml,Higgins:2023gwq}.

%% file: appendices/Appendix_force.tex
\section{B.~~Dark Matter Force Signal}
\label{sec:signal-calc}

\noindent
The total vector force imparted by the ultralight DM field onto an object is given by \cref{eq:dm-force}. For our experiment, the data was gathered for motions along the zenith direction, along which we are most sensitive.

We can then write our sensitivity axis as
\begin{equation}
\begin{split}
    \bvec{\zeta} &\equiv
    (\cos \lambda \cos(\omega_\oplus t + \phi), 
    \cos\lambda \sin(\omega_\oplus t + \phi), 
    \sin \lambda)^\intercal \\
     &\simeq
    (\cos \lambda \cos\phi, 
    \cos \lambda \sin\phi, 
    \sin \lambda)^\intercal
    \,,
\end{split}
\end{equation}
where $\lambda$ and $\phi$ are the latitude and longitude of our experiment, respectively, and $\omega_\oplus$ is the angular frequency of a sidereal day, $\omega_\oplus = 2 \pi / T_\oplus \approx 7.5 \times 10^{-5}\,\mrm{Hz}$. Since our total observation times are always significantly shorter than a day, we make the approximation that $\omega_\oplus t \ll 1$, which we have checked to be appropriate. The relevant force experienced by an object is then 
\begin{equation}
    \begin{split}
    F_\mrm{DM}&(t) \equiv \bvec{\zeta} \cdot \bvec{F}(t)\\ &= g_{B - L} \mathcal{R} m a_0[\alpha_x \cos\lambda \cos\phi \cos(\omega_\mrm{DM}t + \varphi_x) \\
&\hspace{1.875cm}+\,\alpha_y \cos\lambda \sin\phi \cos(\omega_\mrm{DM}t + \varphi_y) \\
&\hspace{1.875cm}+\, \alpha_z \sin\lambda \cos(\omega_\mrm{DM}t + \varphi_z)]\,.
    \end{split}
    \label{eq:dm-force-proj}
\end{equation}

The dynamics of our system can be modelled as two coupled springs in the weak coupling limit. The equations of motion describing the trap and levitated particle along our axis of sensitivity are then, respectively,
\begin{equation}
\begin{split}
\ddot{x}_t + \gamma_t \dot{x}_t + \omega_t^2 x_t  &= \frac{F_{\mathrm{DM}}^t(t)}{m_t}\,, \\
\ddot{x}_p + \gamma_p \dot{x}_p + \omega_p^2 (x_p - x_t) &= \frac{F_{\mathrm{DM}}^p(t)}{m_p}\,,
\end{split}
\label{eq:eoms}
\end{equation}
where $\gamma_i \equiv \omega_i / Q_i$ are the thermal dissipation rates, with $\omega_i$ and $Q_i$ the resonance angular frequencies and quality factors of the respective parts of the setup. We have indicated this via $t$ and $p$, stipulating that we are considering either trap or particle. 
From this system of equations, we can solve for the trap dynamics separately and then use this solution to solve for the levitated particle dynamics. 

For a sinusoidal driving force of the form $F(t) \equiv F_0 \cos(\omega t + \varphi)$, the solution to the driven-damped harmonic oscillator can be written as
\begin{equation}
    x(t) = \lvert \chi(\omega)\rvert F_0 \cos[\omega t + \varphi - \funop{\arg(\funop{\chi(\omega)})}]\,,
\end{equation}
where the transfer function $\chi(\omega)$ is given by
\begin{equation}
    \chi(\omega) \equiv \frac{1}{m(\omega_0^2 - \omega^2 + i\gamma\omega)}\,.
\end{equation}
Assuming the trap to have a resonance frequency far below the DM driving frequency, the trap dynamics follow
\begin{equation}
x_t(t) = - \frac{F_\mathrm{DM}^t(t)}{m_t \omega_\mathrm{DM}^2}\,.
\end{equation}

Inserting this into the equation of motion for the levitated particle, we can read off the total force experienced by it:
\begin{equation}
\begin{split}
F_\mrm{DM}^p(t) = \mathcal{F}[&\alpha_x \cos\lambda \cos\phi \cos(\omega_\mrm{DM}t + \varphi_x) \\
+\, &\alpha_y \cos\lambda \sin\phi \cos(\omega_\mrm{DM}t + \varphi_y) \\
+\, &\alpha_z \sin\lambda \cos(\omega_\mrm{DM}t + \varphi_z)]\,,
\end{split}
\label{eq:force-short-noise}
\end{equation}
where we have defined the quantity
\begin{equation}
\mathcal{F} = g_{B - L} \left(\mathcal{R}_p - \frac{\omega_0^2}{\omega_\mathrm{DM}^2}\mathcal{R}_t \right)  m_p a_0\,.
\label{eq:force-scale-app}
\end{equation}
where we have made the notational replacement $\omega_p \rightarrow \omega_0$.

The total, random force experienced by the levitated particle can be decomposed into the driving dark matter force above and forces stemming from environmental noise,
\begin{equation}
    F_p(t) = F_\mrm{DM}^p(t) + F_\mrm{noise}^p(t)\,,
\end{equation}
We assume that the noise is white, such that it has equal power across all frequencies. For white noise, the force should be normally distributed as $F^p_\mrm{noise} \sim \mathcal{N}(0, \sigma^2)$, where $\sigma^2 \equiv S_{FF}(\omega) \Delta f$ is the force noise variance in the time domain, $S_{FF}(\omega)$ is the force noise power spectral density (PSD), and $\Delta f$ is the sampling frequency. 

We can derive the expected (one-sided) signal PSD using the discrete Fourier transform,
\begin{equation}
    \mathcal P(\omega) \equiv 2 \frac{(\Delta t)^2}{T_\mathrm{obs}} \Bigg| \sum_{n = 0}^{N - 1} F_p(t_n) e^{i \omega n \Delta t} \Bigg |^2\,,
    \label{eq:periodogram-dft}
\end{equation}
where $\omega$ is the angular frequency, $N$ is the number of points sampled in the time domain, and $\Delta t \equiv T_\mathrm{obs} / N$ is the sampling frequency, with $T_\mrm{obs}$ the total observation time. The factor of two accounts for the `folding' of the result from negative angular frequencies to positive angular frequencies to produce the one-sided periodogram. 

From this, we can derive an expression for the signal periodogram normalised by the expected force noise PSD, $S_{FF}$. We call this quantity the excess power, $\Lambda$. Defining the dimensionless parameter
\begin{equation}
    \kappa \equiv \sqrt{\frac{\mathcal{F}^2 T_\mathrm{obs}}{2 S_{FF}}}\,,
    \label{eq:kappa_app}
\end{equation}
we have that
    \begin{align}\label{eq:excess-power-short}
    \Lambda(&\kappa, \bvec{\alpha}, \bvec{\varphi}; \omega) \equiv \frac{\mathcal{P}(\omega)}{S_{FF}(\omega)} \nonumber\\
    &= \kappa^2 [\alpha_x^2 \cos^2 \lambda \cos^2\phi + \alpha_y^2 \cos^2\lambda\sin^2\phi + \alpha_z^2 \sin^2\lambda \nonumber \\
&+ 2 \alpha_x \alpha_y \cos^2\lambda \sin\phi \cos\phi \cos(\varphi_y - \varphi_x) \nonumber \\
&+ 2 \alpha_x \alpha_z \sin\lambda \cos\lambda \cos\phi \cos(\varphi_z - \varphi_x) \\
&+ 2 \alpha_y \alpha_z \sin\lambda \cos\lambda \sin\phi \cos(\varphi_z - \varphi_y)]\,\delta_{\omega,\,\omega_\mrm{DM}}\,, \nonumber
    \end{align}
This expression is similar to that derived in Ref.~\cite{Amaral:2024tjg}; however, their observation time was taken to be longer than a sidereal day, leading to three peaks instead of one.

%% file: appendices/Appendix_materials.tex
\subsection{C.~Neutron-to-Atomic-Weight Ratios}
\label{sec:R-calc}

\noindent
We compute the neutron-to-atomic-weight ratios, $\mathcal{R}$, of the trap and particle by considering the materials composing them. If a composite material is made of component materials $i$, each with total number of neutrons $\mathcal{N}_i$, total number of nucleons $\mathcal{A}_i$, and masses $M_i$, then the total ratio $\mathcal{R}_\mrm{tot}$ can be written as
\begin{equation}
\label{eq:r-tot}
    \mathcal{R}_\mrm{tot} \equiv \frac{\sum_i \mathcal{N}_i}{\sum_i \mathcal{A}_i} = \frac{\sum_i M_i \mathcal{R}_i}{\sum_i M_i}\,,
\end{equation}
where $\mathcal{R}_i \equiv \mathcal{N}_i / \mathcal{A}_i$ are the neutron-to-atomic-weight ratios of each individual material. We summarize our calculations of the total $\mathcal{R}$ values for the particle and trap in \cref{tab:particle-R,tab:trap-R}, respectively.

\begin{table}[t!]
\renewcommand{\arraystretch}{1.5}
    \centering

    \begin{tabular*}{\columnwidth}{@{\extracolsep{\fill}}lcc@{\extracolsep{\fill}}}
        \toprule
        \textbf{Particle \hfill} & &  \hfill\textit{Present} \\

    \toprule
     Material & $M_i$ [mg] & $\mathcal{R}_i$ \\
    
    \midrule 
    $\mrm{Nd_{2} Fe_{14} B}$ (Magnet) & 0.356 & 0.522\\
    $\mrm{Si_{} O_{2}}$ (Glass Bead) & 0.072 & 0.501 \\
    \midrule[0.25pt]
    & $\sum_i M_i = 0.427$& $\mathcal{R}_\mrm{tot} = 0.518$  \\
     \bottomrule  
    \end{tabular*}

\begin{tabular*}{\columnwidth}{@{\extracolsep{\fill}}lcc@{\extracolsep{\fill}}}
        \toprule 
         & &  \hfill\textit{S/M/L}\\

    \toprule

    Material & $M_i$ [mg] & $\mathcal{R}_i$ \\
    \midrule 
    $\mrm{Sm Co_{5}}$ (Magnet) & 0.263 & 0.557  \\
    \midrule[0.25pt]
    & $\sum_i M_i = 0.263$ & $\mathcal{R}_\mrm{tot}= 0.557$  \\

     \bottomrule  
    \end{tabular*}
    \caption{The neutron-to-atomic-weight ratios of the levitated particle in our present and \textit{S}hort-, \textit{M}edium-, and \textit{L}ong-term setups. Shown are the masses of the composing materials, $M_i$, and their neutron-to-atomic-weight ratios, $\mathcal{R}_i$. The total value of this ratio for the composite material, $\mathcal{R}_\mrm{tot}$, is computed via \cref{eq:r-tot}.}

\label{tab:particle-R}
\end{table}

\begin{table}[t!]
\renewcommand{\arraystretch}{1.5}

    \centering

    \begin{tabular*}{\columnwidth}{@{\extracolsep{\fill}}lcc@{\extracolsep{\fill}}}
        \toprule
        \textbf{Trap \hfill} & &  \hfill\textit{Present} \\

    \toprule
     Material & $M_i$ [g] & $\mathcal{R}_i$ \\
    
    \midrule 
    $\mrm{Al}$ (Holder) & 153 & 0.518\\
    $\mrm{Ta}$ (Trap) & 2.94 & 0.597 \\
    $\mrm{PEEK}^\text{\scriptsize{1}}$ (Lid) & 0.100 & 0.480 \\
    $\mrm{Nb}$ (SQUID Shielding) & 12.8 & 0.559\\
    $\mrm{Brass}$ (SQUID Part) & 11.0 & 0.563\\
    $\mrm{Cu}$ (Thermalization) & 0.600 & 0.544 \\
    $\mrm{Metaglass}^\text{\scriptsize{2}}$ (Shielding) & 15.5 & 0.537 \\
    \midrule[0.25pt]
    & $\sum_i M_i = 196$& $\mathcal{R}_\mrm{tot} = 0.526$  \\
     \bottomrule  
    \end{tabular*}

\begin{tabular*}{\columnwidth}{@{\extracolsep{\fill}}lcc@{\extracolsep{\fill}}}
        \toprule 
         & &  \hfill\textit{S/M}\\

    \toprule

    Material & $M_i$ [g] & $\mathcal{R}_i$ \\
    
    \midrule 
    $\mrm{Al}$ (Holder) & 153 & 0.518 \\
    $\mrm{Ta}$ (Trap) & 0.833 & 0.597 \\
    $\mrm{PEEK}^\text{\scriptsize{1}}$ (Lid) & 0.100 & 0.480 \\
    $\mrm{Cu}$ (Thermalization) & 0.600 & 0.544 \\
    \midrule[0.25pt]
    & $\sum_i M_i = 155$& $\mathcal{R}_\mrm{tot} = 0.519$  \\
    \bottomrule  
    \end{tabular*}

\begin{tabular*}{\columnwidth}{@{\extracolsep{\fill}}lcc@{\extracolsep{\fill}}}
        \toprule 
         & &  \hfill\textit{L}\\

    \toprule

    Material & $M_i$ [g] & $\mathcal{R}_i$ \\
    
    \midrule 
    $\mrm{Al}$ (Holder) & 153 & 0.518 \\
    $\mrm{Ta}$ (Trap) & 2.94 & 0.597\\
    $\mrm{PEEK}^\text{\scriptsize{1}}$ (Lid) & 0.100 & 0.480 \\
    $\mrm{Cu}$ (Thermalization) & 0.600 & 0.544 \\
    $\mrm{Al}$ (Container Walls) & 8.11 & 0.518\\
    $\mrm{H}$ (Solid Hydrogen) & 86.0 & 0.008\\
    \midrule[0.25pt]
    & $\sum_i M_i = 253$& $\mathcal{R}_\mrm{tot} = 0.344$  \\
    \bottomrule  
    \end{tabular*}
    \caption{The neutron-to-atomic-weight ratios of the superconducting trap in our present and {$\text{\textit{S}hort-,}$} \textit{M}edium-, and \textit{L}ong-term setups. Shown are the masses of the composing materials, $M_i$, and their neutron-to-atomic-weight ratios, $\mathcal{R}_i$. The total value of this ratio for the composite material, $\mathcal{R}_\mrm{tot}$, is computed via \cref{eq:r-tot}.\newline {$^{\scriptsize{1}}\text{\normalsize Polyether ether ketone }(\mrm{C_{19} H_{12} O_{3}})$.}\newline $^{\scriptsize{2}}\text{\normalsize Based on Metaglass 2714A composition}$.}
    
\label{tab:trap-R}
\end{table}

%% file: appendices/Appendix_stats.tex
\section{D.~Statistical Analysis}
\label{sec:stats}

\subsection{D.1~Derivation of Likelihood Function}

%
%
%

\noindent
For a monochromatic signal such as the one we expect from ULDM, the measured excess power in the bin containing the signal, $p$, should be distributed according to a non-central $\chi^2$ distribution with two degrees of freedom and non-centrality parameter $\Lambda$ given by \cref{eq:excess-power-short}~\cite{groth1975probability,Amaral:2024tjg}. To account for the stochastic ULDM variables, we define the marginalized likelihood
\begin{equation}
    \mathcal{L}(\kappa; p) \equiv \int \funop{\chi^2\mleft(p; 2, \funop{\Lambda(\kappa, \bvec{\alpha}, \bvec{\varphi})}\mright)}\funop{\pi(\bvec{\alpha}, \bvec{\varphi})} \dd^3 \bvec{\alpha} \dd^3 \bvec{\varphi}\,,
    \label{eq:marg-lik}
\end{equation}
where $\pi(\bvec{\alpha}, \bvec{\varphi})$ is the joint probability density of the Rayleigh amplitudes and uniformly distributed phases of the ULDM field. Individually, these random variables follow the probability densities given in \cref{eq:dm-dist}. Since each of them is statistically independent, their joint probability density is the product of their individual densities. We assume that they take constant values over a coherence time, simplifying the theoretical hurdle of modelling their continuous change while still allowing us to account for their randomness in our statistical treatment.


We can perform a change of variables from the above polar space to a Cartesian one. The random amplitudes and phases can then be described by the three random vectors $\bvec{x} \equiv (x_1, x_2)^\intercal$, $\bvec{y} \equiv (y_1, y_2)^\intercal$, and $\bvec{z} \equiv (z_1, z_2)^\intercal$. Each component of $\bvec{x}$, $\bvec{y}$, and $\bvec{z}$ then has a mean of zero, with variances of $(\cos^2\lambda \cos^2\phi)/2$, $(\cos^2\lambda \sin^2\phi)/2$, and $(\sin^2\lambda)/2$, respectively. In this basis, the non-centrality parameter can be compactly written as
\begin{equation}
    \Lambda(\kappa) = \kappa^2 |\bvec{x} + \bvec{y} + \bvec{z}|^2.
\end{equation}
It is convenient to perform one further change of variables to $\bvec{u} \equiv \bvec{x} + \bvec{y} + \bvec{z}$, $\bvec{v} \equiv \bvec{x} + \bvec{y} - \bvec{z}$, and $\bvec{w} \equiv \bvec{x} - \bvec{y} - \bvec{z}$. The last two definitions are required to complete our transformation. Each component of these vectors is now normally distributed with zero mean and variance $1/2$. The joint probability density then transforms as $\pi(\bvec{\alpha}, \bvec{\varphi}) \dd^3 \bvec{\alpha} \dd^3 \bvec{\varphi} \rightarrow \pi'(\bvec{u}, \bvec{v}, \bvec{w}) \dd^2 \bvec{u} \dd^2 \bvec{v} \dd^2\bvec{w}$, with
\begin{equation}
\pi'(\bvec{u}, \bvec{v}, \bvec{w}) \\
    = \frac{1}{\pi^3} e^{-|\bvec{u}|^2} e^{-|\bvec{v}|^2} e^{-|\bvec{w}|^2}\,.
\end{equation}

In this basis, we can expand the non-central $\chi^2$ distribution in terms of the modified Bessel function, $I_0$, to write the likelihood in \cref{eq:marg-lik} as
\begin{equation}
        \mathcal{L}(\kappa; p) = \frac{e^{- p / 2}}{2}\sum_{n = 0}^{\infty}\frac{p^n}{(2^n n!)^2} \funop{\mathcal{I}_n(\kappa)}\,,
        \label{eq:lik-bessel}
\end{equation}
with the integral $\mathcal{I}_n(\kappa)$ given by
\begin{equation}
    \mathcal{I}_n(\kappa) \equiv \int e^{-\frac{\kappa^2 |\bvec{u}|^2}{2}}  (\kappa \, |\bvec{u}|)^{2 n} \funop{\pi'} \dd^2 \bvec{u} \dd^2 \bvec{v} \dd^2\bvec{w}\,.
\end{equation}
We can analytically evaluate this using the integral definition of the gamma function, yielding
\begin{equation}
    \mathcal{I}_n(\kappa) = \frac{2 \kappa^{2n} (2^n n!) }{(2 + \kappa^2)^{n+1}}\,.
\end{equation}
Substituting this into \cref{eq:lik-bessel}, we arrive at our result:
\begin{equation}
    \mathcal{L}(\kappa; p) = \frac{1}{2 + \kappa^2} e^{-p / (2 + \kappa^2)}\,.
    \label{eq:lik-coh}
\end{equation}

\begin{figure}
    \centering
    \includegraphics{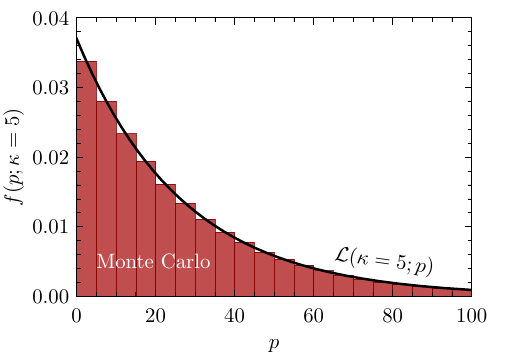}
    \caption{Distribution of the measured excess power $p$ in the coherent regime for $\kappa = 5$ (see \cref{eq:kappa_app}). Data was simulated from $10^6$ Monte Carlo runs beginning from \cref{eq:force-short-noise}. The solid line shows the analytical result of \cref{eq:lik-coh} and is in excellent agreement with simulations.}
    \label{fig:low-obs-stats}
\end{figure}

This likelihood suggests that the statistics of our problem do not depend on the positional parameters $\lambda$ and $\phi$. This is because our observation times are shorter than the sidereal period of the Earth, such that we treat our detector as being effectively stationary. Since the ULDM field is isotropic, with each random variable being identically and independently distributed, it should not matter where on Earth we are when making inferences. Thus, while the measured force and periodogram do depend on our position, statistical conclusions do not.

We verify this likelihood using Monte Carlo simulations. We generate $10^6$ force signals according to \cref{eq:force-short-noise} in the time domain, arbitrarily taking the DM angular frequency to be $\omega_\mrm{DM} = 2\pi\,\unit{\hertz}$, with $\mathcal{F} = \SI{5}{\newton}$, $T_\mrm{obs} = \SI{10}{\s}$, $\lambda = 45^\circ$, $\phi = 30^\circ$, and $S_{FF} = \SI{5}{\newton\squared\per\hertz}$, giving $\kappa = 5$. In each simulation, we draw $\alpha$ and $\varphi$ from the distributions shown in \cref{eq:dm-dist}. Finally, we take the one-sided PSD using the discrete Fourier transform in \cref{eq:periodogram-dft}. We show the distribution of the excess power $p$ in the signal-containing bin in \cref{fig:low-obs-stats}. We find excellent agreement with our derived likelihood, \cref{eq:lik-coh}. 


\subsection{D.2~Inferencing Framework}

\noindent
To perform our inferences, we use the likelihood ratio
\begin{equation}
    \tilde{\lambda}(\kappa) \equiv 
\begin{cases}
\frac{\mathcal{L}(\kappa;\,p)}{\mathcal{L}(\hat{\kappa};\,p)} \quad &\text{if} \quad \hat{\kappa} \geq 0\,, \\[2.5ex]
\frac{\mathcal{L}(\kappa; p)}{\mathcal{L}(0; p)}  \quad &\text{else}\,.
\end{cases}
\label{eq:lik-ratio}
\end{equation}
The parameter $\hat{\kappa}$ is the maximum-likelihood estimator of $\kappa$, maximising the likelihood given the measured value of the excess power $p$. The piecewise condition captures the physicality of the parameter $\kappa$: if $\hat{\kappa}$ falls outside of its physical domain of $\kappa \in [0, \infty)$,  $\hat{\kappa}$ is set to $0$. We then construct the two-sided test statistic (TS)~\cite{Cowan:2010js}
\begin{equation}
    \tilde{t}_\kappa \equiv -2 \ln \tilde{\lambda}(\kappa)\,,
    \label{eq:test-stat}
\end{equation}
which is used to assess whether the null hypothesis, defined according to the likelihood in the numerator of \cref{eq:lik-ratio}, is a good description of the observed data compared to the alternative hypothesis, defined as per the maximised unconditional likelihood. A value of $0$ for $\tilde{t}_\kappa$ indicates perfect agreement with the null hypothesis, with larger values indicating greater disagreement. 

To reject the null hypothesis, we define a value for the test statistic that is deemed too extreme to have come from that model.
Specifically, if $f(\tilde{t}_\kappa|\kappa)$ is the distribution of $\tilde{t}_\kappa$ values that we would expect to arise from the null hypothesis defined by the parameter $\kappa$, then we can define the $p$-value of the observed $\tilde{t}_\kappa$,  $\tilde{t}_\kappa^\mrm{obs}$, as
\begin{equation}
    p_\kappa \equiv \int_{\tilde{t}_\kappa^\mrm{obs}}^{\infty} f(\tilde{t}_\kappa | \kappa) \dd \tilde{t}_\kappa\,,
\end{equation}
telling us how probable it is for us to have observed a value of $\tilde{t}_\kappa$ at least as extreme as $\tilde{t}_\kappa^\mrm{obs}$. We find $f(\tilde{t}_\kappa|\kappa)$ from a series of Monte Carlo (MC) simulations, generating $10^6$ pseudo-datasets distributed according to the likelihood given in \cref{eq:lik-coh} for each value of $\kappa$ begin tested. For each pseudo-dataset, we then compute the value of $\tilde{t}_\kappa$ from \cref{eq:test-stat}. The distribution of these values is $f(\tilde{t}_\kappa|\kappa)$.

To assess the discovery significance of our data, we consider $p_0$. We first compute the value of the excess power $p = \mathcal{P} / S_{FF}$ within every frequency bin, where $\mathcal{P}$ and $S_{FF}$ are the data and fit given in the main text, respectively. We then compute the value of $\tilde{t}_0$ within each bin. We find our lowest local value to be $p_0 \approx 2.0 \times 10^{-2}$, an order of magnitude greater our preset condition for discovery, set at the \textit{local} $p$-value of $p_0 \approx 2.7 \times 10^{-3}$ for a $3\sigma$ significance. We therefore cannot claim a discovery and turn to the task of limit setting.

\begin{figure}[t]
    \centering
    \includegraphics{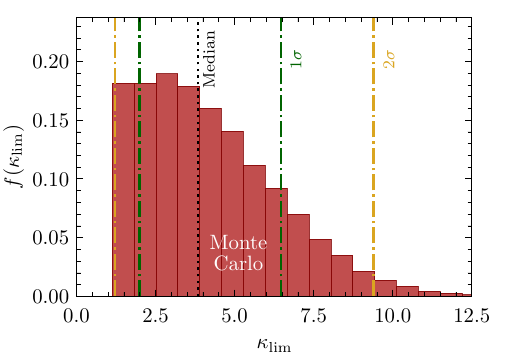}
    \caption{Distribtion of $90\%$ confidence level limits on $\kappa$ derived from our Monte Carlo procedure. The median value is given by $\kappa_\mrm{lim}^\mrm{med} \approx 3.85$, and the $1\sigma$ and $2\sigma$ bands are given by $\kappa_\mrm{lim}^{1\sigma} \in [2.01, 6.47]$ and $\kappa_\mrm{lim}^{2\sigma} \in [1.23, 9.39]$, respectively.}
    \label{fig:beta-lim-dist}
\end{figure}

We proceed similarly to compute our $90\%$ confidence level (CL) limits. We define the false positive rate as $\alpha = 0.1$ and find that value of $\kappa$, $\kappa_\mrm{lim}$, for which $p_\kappa = \alpha$ within each frequency bin. We use our measured PSD for our data-driven limit, whereas we assume that we observe background-only data for our expected upper limits. For the latter, we generate $10^6$ background-only pseudo-datasets according to \cref{eq:lik-coh} with $\kappa = 0$. For each of these datasets, we find $\kappa_\mrm{lim}$, producing a distribution of $\kappa_\mrm{lim}$ values, shown in \cref{fig:beta-lim-dist}. We derive a median value of $\kappa_\mrm{lim}^\mrm{med} \approx 3.85$, with $1\sigma$ and $2\sigma$ bands respectively given by $\kappa_\mrm{lim}^{1\sigma} \in [2.01, 6.47]$ and $\kappa_\mrm{lim}^{1\sigma} \in [1.23, 9.39]$. To find the corresponding quantities for $g_{B - L}$, we use \cref{eq:force-scale-app}. While the two-sided TS construction in \cref{eq:test-stat} generally leads to two-sided limits, we only show the upper limits since we cannot claim discovery.

%% file: appendices/Appendix_future_experiments.tex
\section{E.~Future Experiments}

\noindent
We have three main sources of force noise: external vibrations, thermal noise from the environment that is at some equilibrium temperature and to which the particle is coupled through the dissipation of its mechanical motion, and backaction noise arising from the measurement of the particle position. Currently, we are limited by external vibrations, and therefore any improvements in vibration isolation will improve our force sensitivity. At our present resonance frequency, we have vibration isolation at the level of \SI{80}{\decibel}. At the current $Q$ factor of the resonator, we will need to improve this isolation by roughly \SI{20}{\decibel} to reach thermal motion at \SI{50}{\milli\K}. As the $Q$ factor increases or temperature decreases, the vibration isolation will need to be improved as $(Q / T)^{1/2}$, equivalent to \SI{10}{\decibel} for every factor order of magnitude increase in $Q$ or decrease in temperature. Moreover, for every order of magnitude increase in the particle mass, we will need to improve this isolation by \SI{20}{\decibel}. We note that vibration isolation typically works better at higher frequencies.

For the backaction noise, the position of the magnetic particle is first converted to a flux in a pick-up coil, and this flux is subsequently measured with a SQUID. The backaction noise then arises from the current noise in the SQUID, which couples back as a current noise in the pick-up coil, exerting a force on the particle. The parameter needed to quantify the amount of backaction noise is the energy coupling, $\beta^2 \equiv L_\text{tot} I^2 / (m_p v_p^2)$, which is the ratio between the induction energy produced by the currents in the superconducting circuit due to the motion of the particle and the kinetic energy of the particle. The backaction force is then determined by the SQUID noise, $S_{\Phi\Phi}$. At a given energy coupling, SQUID noise, and $Q$ factor, we can calculate the minimum temperature $T$ of the experiment below which the backaction of the SQUID becomes dominant over the thermal noise of the experiment. This occurs when 
\begin{align}
    4 k_B T \frac{m_p \omega_0}{Q} 
    &> k \beta^2 \frac{L_\text{tot}}{k_\text{M}^2 L_\text{inp}} 2 n_\text{SQ} \hbar\,.
\end{align}
Here, $m_p$ is the mass of the particle, $\omega_0$ is its resonance frequency in the trap, $Q$ is the quality factor, $k$ is the spring constant, $L_\text{tot}$ is the total inductance of the superconducting circuit, $k_\text{M}$ is the mutual inductance coupling constant, $L_\text{inp}$ is the SQUID input inductance, and $n_\text{SQ}$ is the energy resolution of the SQUID. This energy resolution is given by $S_{\Phi\Phi} / (2 L_\text{SQ}) = n_\text{SQ} \hbar$.

Force sensitivity is not the only thing to consider. To exclude ULDM over a larger frequency range, we must increase the bandwidth over which a resonator is sensitive.  
The detection noise has an important effect on the bandwidth of the experiment. Rather than the familiar `full-width half-maximum' bandwidth, $\Delta f_\mrm{FWHM} \equiv f_0/Q$, the detectable bandwidth is used. This is the bandwidth for which the position noise due to the fluctuations in the motion of the resonator is larger than the incorporated detection noise. This occurs at the bandwidth
\begin{equation}
    \Delta f_\text{det} \equiv \frac{f_\mrm{0}}{Q}\sqrt{R_S-1}\,,
\end{equation}
where $R_S \equiv S_{xx,\,\mrm{motion}} /  S_{xx,\,\mrm{SQ}} = S_{FF} \cdot |\chi(\omega_0)|^2 /  S_{xx,\,\mrm{SQ}}$ is the ratio between the two contributions to the position noise. This relates the detectable bandwidth to the energy coupling, quality factor, temperature, and energy resolution of the SQUID. 
In the limit that the backaction noise dominates over the force noise, $R=Q^2 \beta^4$ and the bandwidth reduces to
\begin{equation}
    \Delta f_\text{opt} = f_\mrm{0} \beta^2_\mrm{opt}.
    \label{eq:Bandwidth}
\end{equation}

These factors, together with the particle mass, number of particles, and the difference between the neutron-to-atomic-weight ratios of the levitated particle and trap, are factors that dictate the short-, medium-, and long-term upgrades for our experiment.

When searching for ULDM at an unknown frequency, we would like to stress the importance of the bandwidth, which as implied by \cref{eq:Bandwidth} grows with coupling $\beta^2$. If we already the frequency at which to search for ULDM, we would choose the resonator's resonance frequency to coincide with the frequency of the ULDM. Moreover, the bandwidth of the experiment would not be as important as long as it was larger than the bandwidth necessary to detect the time-varying amplitude and phase of the ULDM field. If the time scale at which amplitude and phase vary is of order $10^6$ Compton cycles, then a bandwidth of the $f_\mrm{0}/10^6$ would suffice. This can be achieved using a coupling value of $\beta^2 \sim 10^{-6}$. 

However, as long as the frequency is unknown, the optimal bandwidth depends on whether there is enough time available to spend $10^6$ cycles at each frequency. If one wants to cover a large frequency in a limited amount of time, one could opt for a coupling larger than the optimal coupling, such as to have a larger bandwidth for each single experiment. This, however, comes at the expense of a larger backaction noise. 
On the one hand, this has the advantage that we need to do fewer experiments since, for a larger bandwidth per experiment, it takes fewer experiments to cover a full bandwidth. This comes at a cost because each of these experiments has a larger noise.

If the time to cover a full bandwidth is limited, the higher backaction noise is cancelled by the longer measurement time per frequency, and the limits are independent of $Q$ and $T$. However if we can measure for a long time, such that one can wait for a set amount of cycles at each frequency, the noise is lowest for lower temperatures and higher $Q$ factors.
In our projections, we decided on a total measurement time of two years with optimal coupling and $4.05 \times 10^5$ cycles at each frequency.

\begin{center}
\begin{figure}[b!]
    \centering
\includegraphics{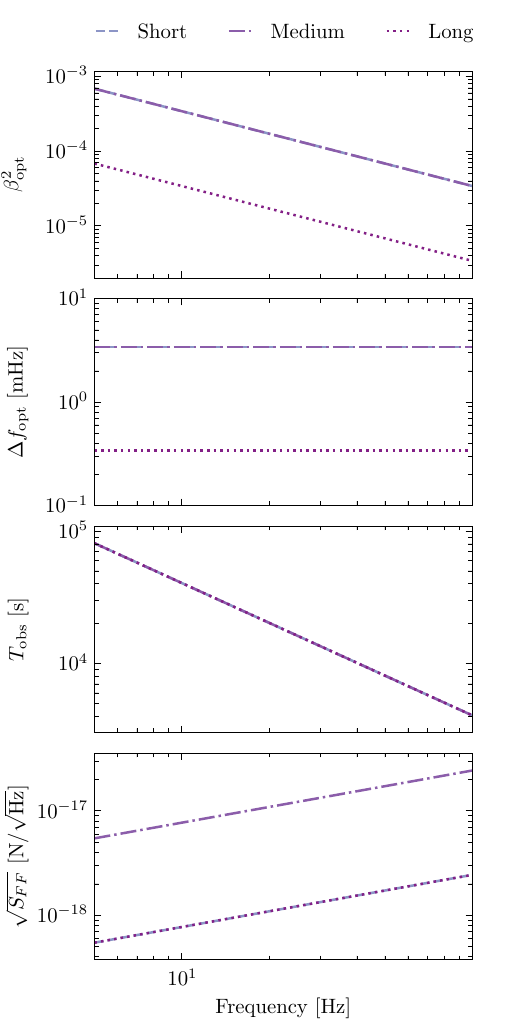}
    \caption{
Quantities needed to compute our projected limits. These are the optimal SQUID coupling ($\beta^2_\mrm{opt}$), the optimal bandwidth that gives us equal thermal and backaction noise ($\Delta f_\mrm{opt}$), the observation time for $4.05 \times 10^5$ Compton cycles ($T_\mrm{obs}$), and the total root force noise PSD ($\sqrt{S_{FF}}$). The relevant parameters needed to compute these quantities are given in the main text.}
    \label{fig:Quantities}
\end{figure}
\end{center}

In \cref{fig:Quantities}, we plot the relevant quantities that we aim for in the short-, medium- and long-term scenarios. We believe the optimal $\beta^2$ can be reached since it is only a factor $10$ to $100$ larger than the $\beta^2$ in the realized experiment we analyzed. The detectable bandwidth then follows from $\beta^2$. The force noise follows from the proposed parameters of our future upgrades. The force noise is approximately $40\,\mrm{dB}$ lower than the data presented. We believe this is achievable because we are currently limited by our vibration isolation system.

\vfill
\clearpage